\documentclass[aps,prl,showpacs,twocolumn,superscriptaddress,amsmath,amssymb]{revtex4-1}

\usepackage{floatrow}
\usepackage{float}
\usepackage[caption=false]{subfig}
\usepackage{graphicx}% Include figure files
\usepackage{dcolumn}% Align table columns on decimal point
\usepackage{bm}% bold math

\usepackage{color}
\usepackage{amssymb}
\usepackage{epstopdf}

\begin{document}

\title{Engineering excited-state interactions at ultracold temperatures}

%% Notice placement of commas and superscripts and use of &
%% in the author list

\author{Michael Mills}
\author{Prateek Puri}
\affiliation{Department of Physics and Astronomy, University of California, Los Angeles, California 90095, USA}	
\author{Ming Li}
\affiliation{Department of Physics, Temple University, Philadelphia, Pennsylvania 19122, USA}
\author{Steven J. Schowalter}
\author{Alexander Dunning}
\author{Christian Schneider}
\affiliation{Department of Physics and Astronomy, University of California, Los Angeles, California 90095, USA}
\author{Svetlana Kotochigova}
\affiliation{Department of Physics, Temple University, Philadelphia, Pennsylvania 19122, USA}
\author{Eric R. Hudson}
\affiliation{Department of Physics and Astronomy, University of California, Los Angeles, California 90095, USA}
 
\date{\today}

\begin{abstract}
Using a recently developed method for precisely controlling collision energy, we observe a dramatic suppression of inelastic collisions between an atom and ion (Ca + Yb$^+$) at low collision energy.
This suppression, which is expected to be a universal phenomenon, arises when the spontaneous emission lifetime of the excited state is comparable to or shorter than the collision complex lifetime.
We develop a technique to remove this suppression and engineer excited-state interactions.
By dressing the system with a strong catalyst laser, a significant fraction of the collision complexes can be excited at a specified internuclear separation.
This technique allows excited-state collisions to be studied, even at ultracold temperature, and provides a general method for engineering ultracold excited-state interactions.
\end{abstract}
\maketitle

In the last quarter century, the development of techniques for producing ultracold matter led to important advances in the understanding of fundamental collision physics. 
The ability to observe few and even single partial wave collision events allowed the observation of quantum threshold behavior and unitarity limited processes~\cite{Thomas2004,Ospelkaus2010,Ni2010}.
It also revealed the impact chemical binding forces, quantum statistics, internal structure, and dimensionality have on collisions, as well as provided the potential for control of chemical reactions~\cite{Jin99,Jones2006,Lang2008,Hudson2006,Ratschbacher2012,TG}. 

The overwhelming majority of these studies were performed with collision partners in their ground electronic state. 
This is at least partially due to the fact that interactions at ultracold temperatures tend to naturally suppress electronically excited collisions, as pointed out in \cite{Julienne89} and demonstrated in \cite{Gould95a,BaCl}. 
This suppression arises as the long-range interactions between collision partners tend to shift any laser that would electronically excite one of the collision partners out of resonance at very long range.
Therefore, in order for an electronically excited collision to occur, the atom or molecule must remain in the excited state until the collision pair reaches the short range where inelastic collisions and chemistry can occur. 
The combination of the extremely low energy of ultracold matter with the typically short lifetime of an electronically excited state means that most excited collision partners have decayed back to their ground state by the time they are close enough to interact. 
This effect, dubbed reaction blockading~\cite{BaCl}, is especially strong in systems with long-range interactions such as atom-ion or molecule-molecule pairs. 

Collisions involving electronically excited atoms and molecules play an important role in processes such as combustion~\cite{combustion}, explosives, atmospheric chemistry~\cite{atmosphericChemistry}, stellar evolution~\cite{stellarEvolution}, interactions in the interstellar medium~\cite{interstellarMedium}, and the formation of new molecules~\cite{Puri2017}, yet many studies of these reactions have been limited to high temperature, where quantum effects are often obscured.
As such, it is highly desirable to find a general technique to enable the study of such collisions at ultracold temperatures, allowing the precision and control afforded by the ultracold regime to be brought to bear on these important problems. 
Here, we demonstrate such a technique in a prototypical atom-ion system. 
Building from work on hyperfine-changing collisions in laser-cooled systems~\cite{Walker89,Gould95a,Gould95}, we apply a strong laser field that dresses the system and promotes the molecular collision complex to a specified excited state at a specified range. 
In this way, we engineer the electronic excitation of collision partners at short range and extract the excited channel rate constants.
Additionally, by controlling the range at which this laser addresses the reactants, this technique is sensitive to the features of molecular potentials, enabling a new class of experiments to probe molecular potentials at controlled internuclear separations.  
Excitingly, the technique appears to be completely general and can be applied at higher temperatures.

In what follows, we use a recently developed method~\cite{Shuttling} for precisely controlling collision energy to study the charge-exchange collision between Ca(4s4p $^1$P$_1$) + Yb$^+$(6s $^2$S$_{1/2}$) as a function of collision energy from 0.05~K to 0.65~K. 
From this data, we observe reaction blockading of the charge-exchange rate and measure the dependence of this suppression on collision energy. 
Finally, we introduce a strong laser to dress the system and observe an increased charge-exchange rate for the Ca(4s4p $^1$P$_1$) + Yb$^+$(6s $^2$S$_{1/2}$) channel, effectively eliminating the suppression.
A quantum coupled-channels calculation is presented and shows good agreement with the data.
The technique is qualitatively explained using a simple semi-classical model based on dressed molecular potentials and a Landau-Zener type transition.

\floatsetup[figure]{subcapbesideposition=top}
\begin{figure}
 \centering
 \sidesubfloat[c]{\includegraphics[width=.86\textwidth]{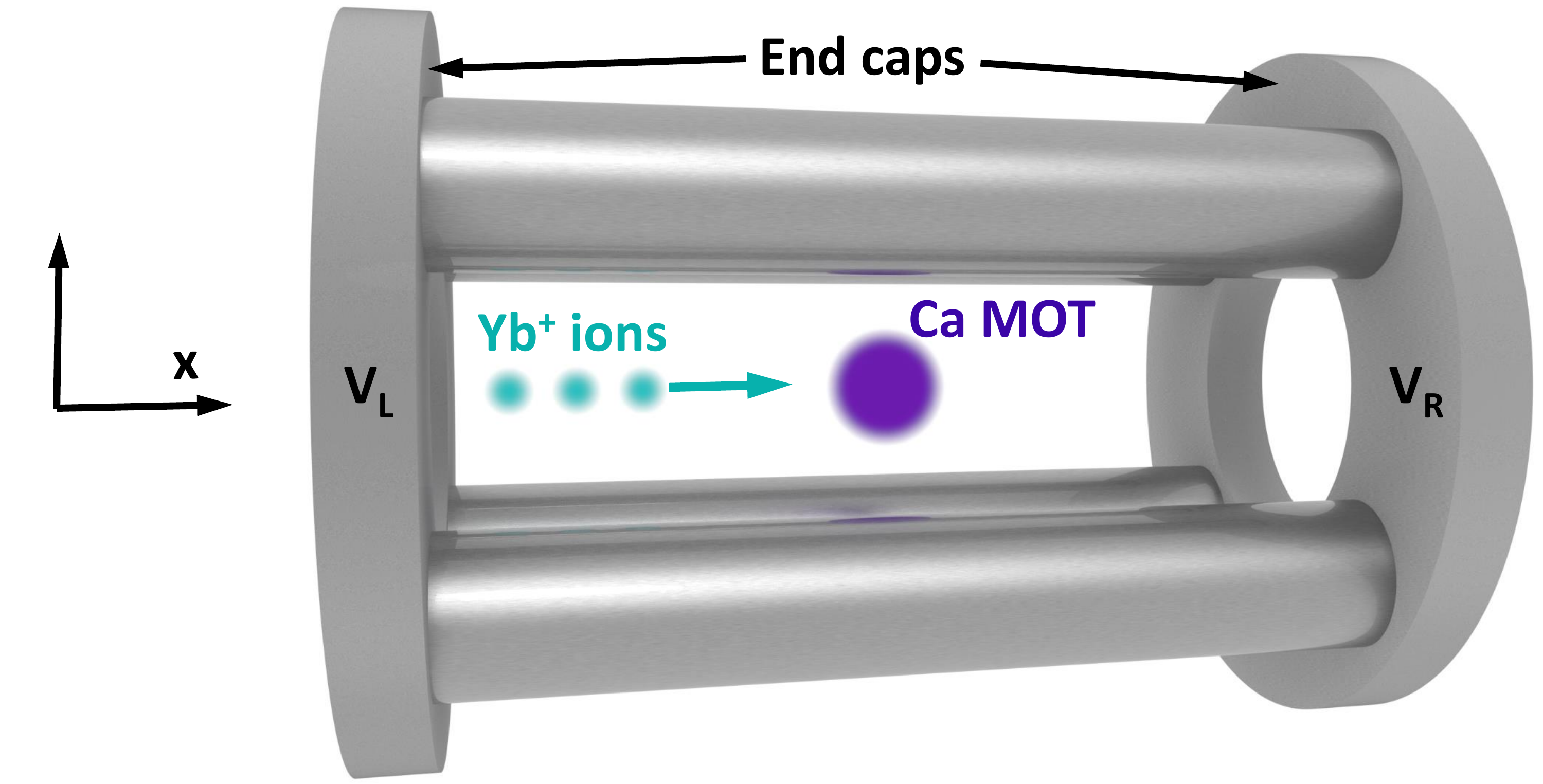}}
 \hfill
 \vspace{6pt}
 \sidesubfloat[c]{\includegraphics[width=.93\textwidth]{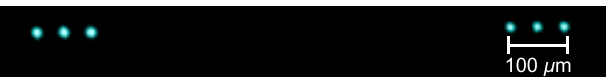}}
 \hfill
 \vspace{6pt}
 \sidesubfloat[c]{\includegraphics[width=.93\textwidth]{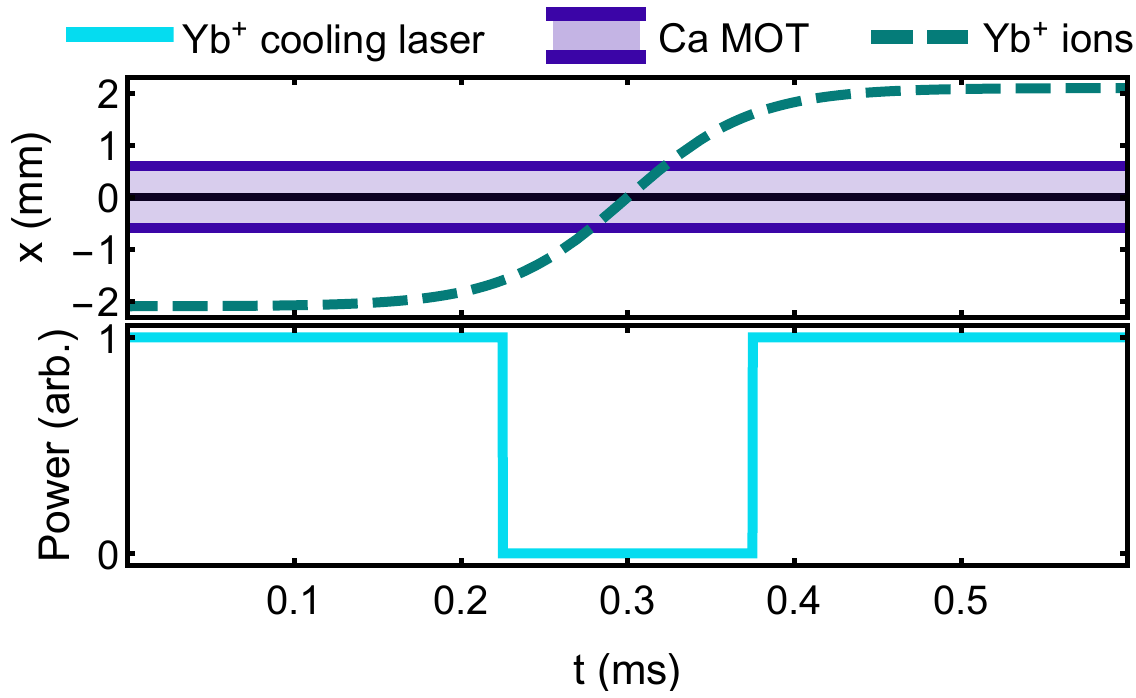}}
 \hfill
 \vspace{6pt}
 \sidesubfloat[]{\includegraphics[width=.95\textwidth]{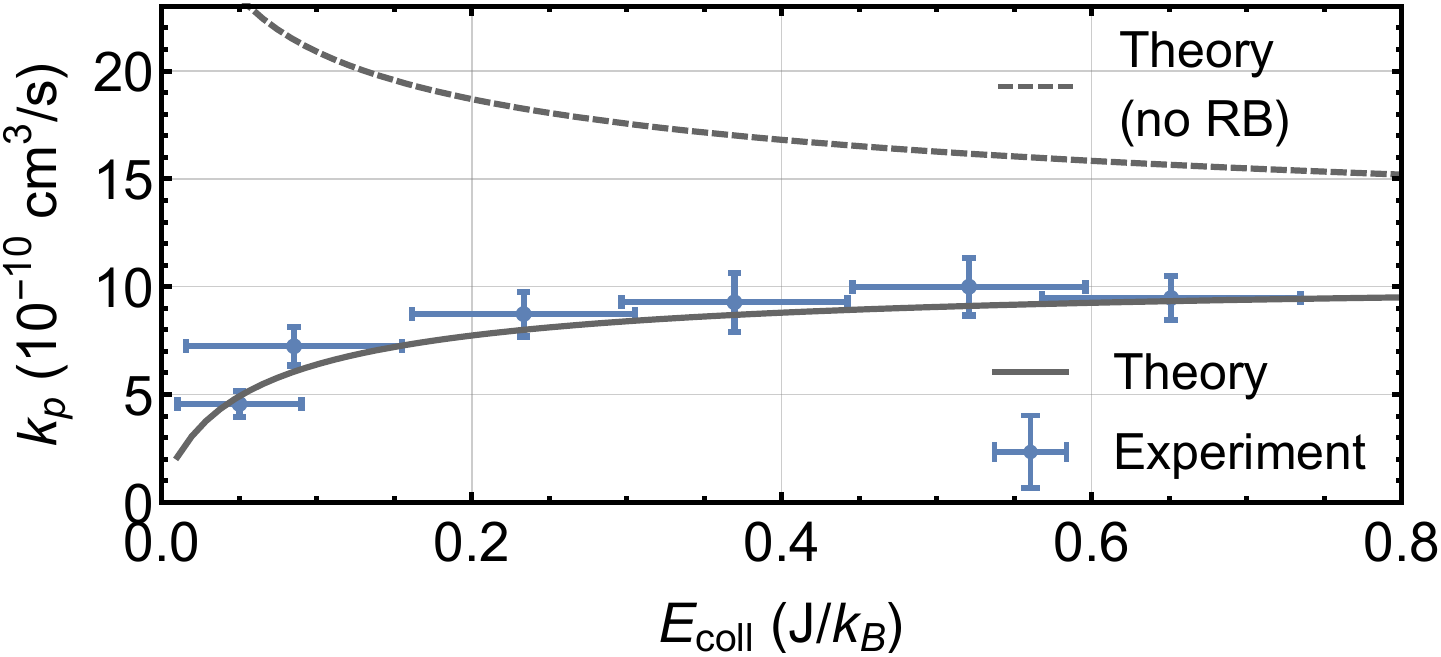}}
 \hfill
 \caption{\textbf{Shuttling in the hybrid atom-ion MOTion trap.} \textbf{(a)} Simplified illustration (not to scale) of the MOTion trap. By smoothly ramping the voltages $V_L$ and $V_R$ on the ion trap endcaps, the Yb$^+$ ions can be shuttled through the Ca MOT at a controlled collision velocity. 
\textbf{(b)} 
Long-exposure false-color fluorescence image of 3 shuttled Yb$^+$ ions. As the exposure time is much greater than the shuttling period, fluorescence from the 3 ions is concentrated at the positions of the two end points on either side of the MOT (located in the center of this image), where the ions spend the most time.
\textbf{(c)} 
Experimental sequence as a function of time $t$ illustrating the shuttling technique. 
As the Yb$^+$ ions are shuttled through the Ca MOT, the 369 nm Yb$^+$ cooling beams are extinguished to prepare the ions in the 6s $^2$S$_{1/2}$ state. 
\textbf{(d)} 
Measured charge-exchange rate coefficient for Ca($^1$P$_1$) + Yb$^+$($^2$S$_{1/2}$) as a function of collision energy using the shuttling technique. 
Also shown are rate coefficients from coupled-channels calculations, one with (solid line) and one without (dashed line) the effect of reaction blockading (RB). 
The decrease in rate at low temperature is indicative of reaction blockading of collisions along the excited atomic entrance channel. 
Error bars correspond to the standard error in experimental measurements.
}\label{fig:trap}
\end{figure}

The experiment is performed in the second-generation \mbox{MOTion} trap, sketched in Fig.~\ref{fig:trap} and described in detail in Refs.~\cite{Rellergert2013,Schowalter2016}. 
Coulomb crystals of Yb$^+$ ions are confined in a segmented radio-frequency linear quadrupole trap, while a small Ca oven provides atomic Ca vapor which is loaded into a magneto-optical trap (MOT). 
The atoms, ions, and their overlap are detected by imaging onto EMCCD cameras. 
By smoothly ramping the voltages of the ion trap end caps, we can shuttle an ion chain through the MOT at a controlled velocity, illustrated in Fig.~\ref{fig:trap}(a), (b), and (c) and described in detail in \cite{Shuttling}, allowing precise control of the reactant collision energy.
By extinguishing the 369 nm Yb$^+$ cooling laser when the Yb$^+$ ions are shuttled through the Ca MOT, the ions are prepared in the ground 6s~$^2$S$_{1/2}$ state.
Using this shuttling method, we measure the charge-exchange rate of Ca(4s4p $^1$P$_1$) + Yb$^+$(6s $^2$S$_{1/2}$) as a function of collision energy and observe reaction blockading of the rate, shown in Fig.~\ref{fig:trap}(d).
Specifically, for an ion chain with 50 mK collision energy (defined as $\langle E_{col} \rangle / k_B$, where $\langle E_{col} \rangle$ is the average kinetic energy in the center of mass frame and $k_B$ is the Boltzmann constant), we measure a rate constant of $k_p = (4.6 \pm 0.6)\times 10^{-10}$ cm$^3$/s, compared to the no-suppression theoretical prediction of $k_p = 23 \times 10^{-10}$~cm$^3$/s, an observed suppression factor of $\sim5$.

This reaction blockading can be understood by considering the long-range atom-ion interaction \cite{Julienne89,BaCl}. At long range the atom and ion interact primarily through the charge-induced dipole and charge-quadrupole potentials of the forms $-\frac{\alpha}{2}R^{-4}$ and $-\frac{Q}{2} (3\cos^2(\theta)-1) R^{-3}$, respectively, where $R$ is the atom-ion separation, $\alpha$ is the neutral atom polarizability, $Q$ is the neutral atom quadrupole moment, and $\theta$ is the angle between the quadrupole moment and the internuclear axis.
Thus, a laser resonant with two atomic states at long range, which have different polarizabilities and quadrupole moments, is no longer resonant when the atom and ion are in close proximity.
For the Ca $^1$P$_1\leftarrow$ $^1$S$_0$ transition with linewidth $\Gamma$ and a laser detuning $\delta = -\Gamma = 2\pi \times$($-34.6$~MHz), the laser becomes resonant at $R \approx 1300$ $a_0$ and becomes detuned by $10 \Gamma$ at $\sim 600$ $a_0$. 
Therefore, for a charge-exchange event to occur, the atom-ion pair must propagate inward without the Ca atom decaying from this distance to distances of $\sim 40$ $a_0$, where couplings to other potentials become significant. 
For collision temperatures greater than $\gtrsim 10$~K, the atom-ion pair approaches quickly enough such that the Ca $^1P_1$ state is unlikely to decay before reaching short range, affecting the rate coefficient by $\lesssim 1\%$. 
For a collision temperature of $1$ mK, however, this effect leads to a suppression by a factor of $\sim$100.

\begin{figure*}[t]
 \centering

\sidesubfloat[]{\includegraphics[width=.32\textwidth]{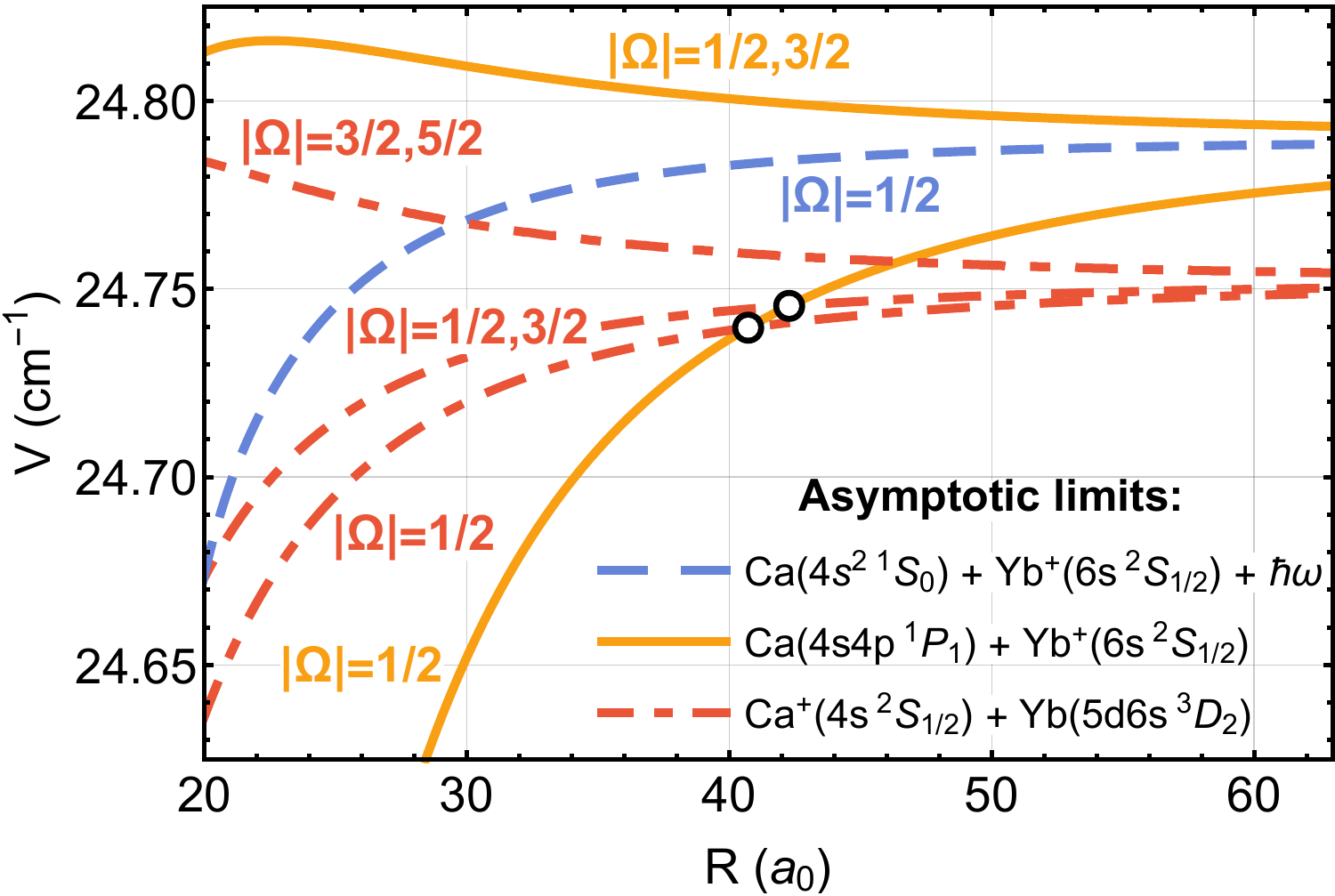}}
\sidesubfloat[]{\includegraphics[width=.185\textwidth]{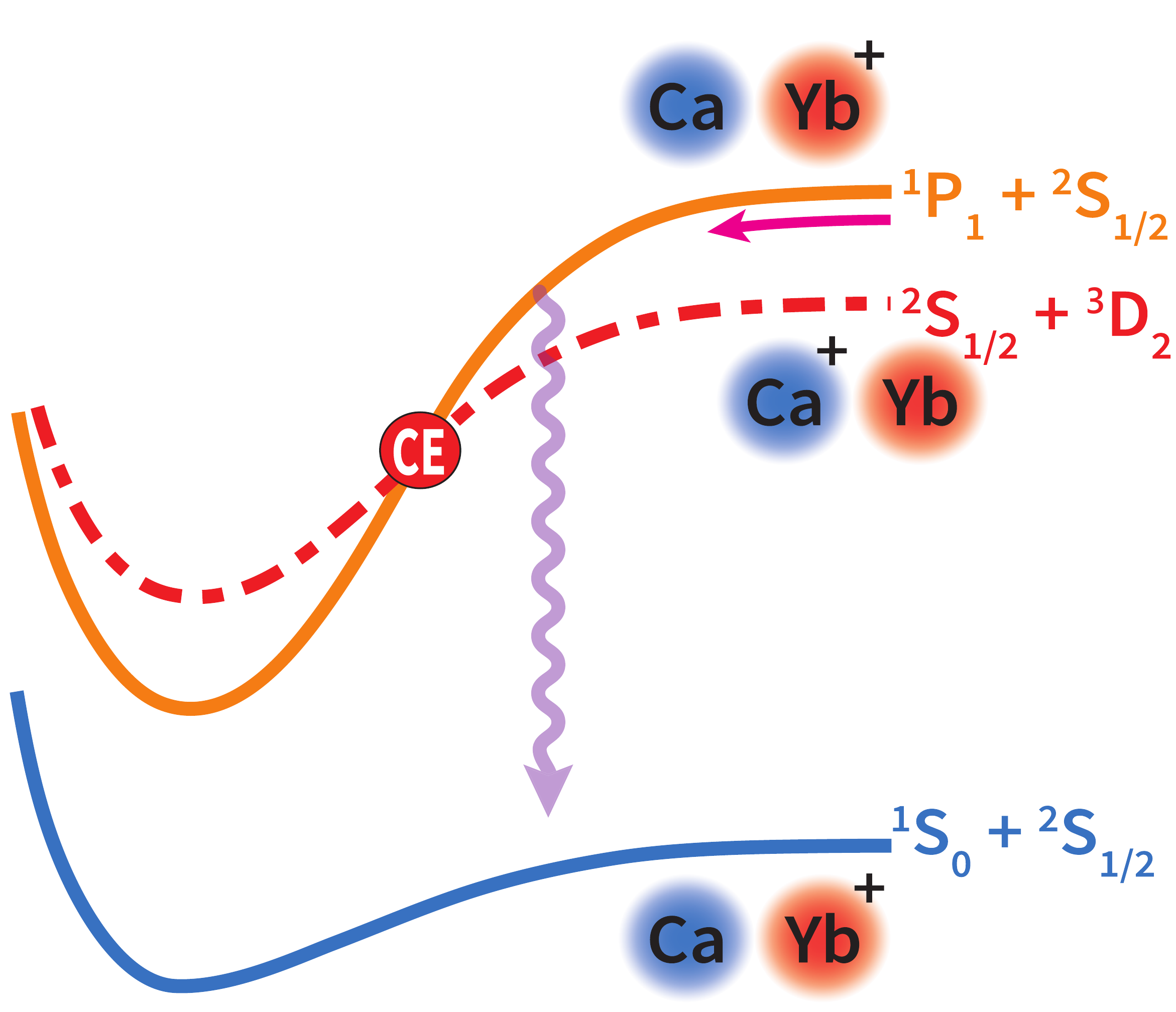}}
\sidesubfloat[]{\includegraphics[width=.185\textwidth]{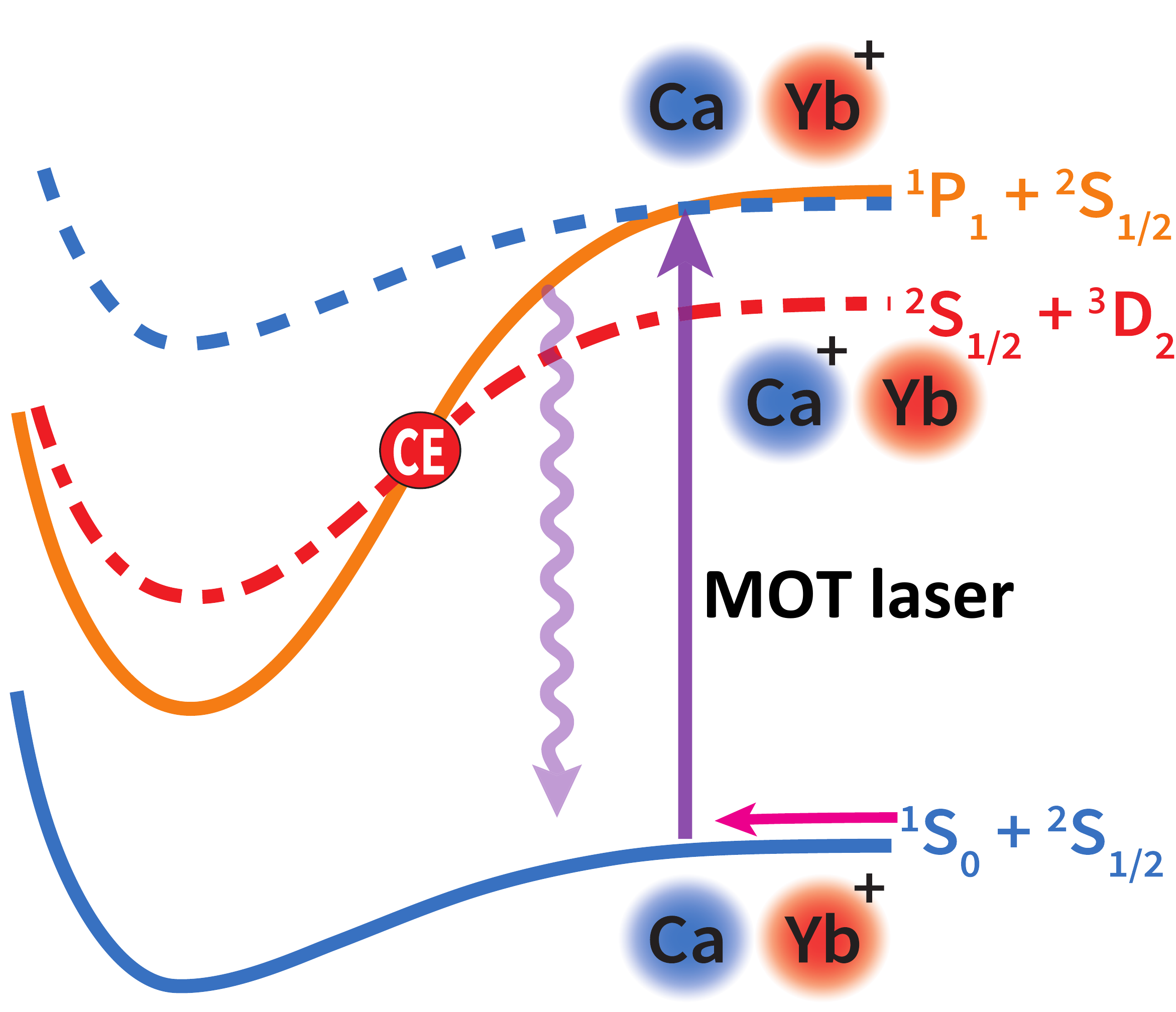}}
\sidesubfloat[]{\includegraphics[width=.185\textwidth]{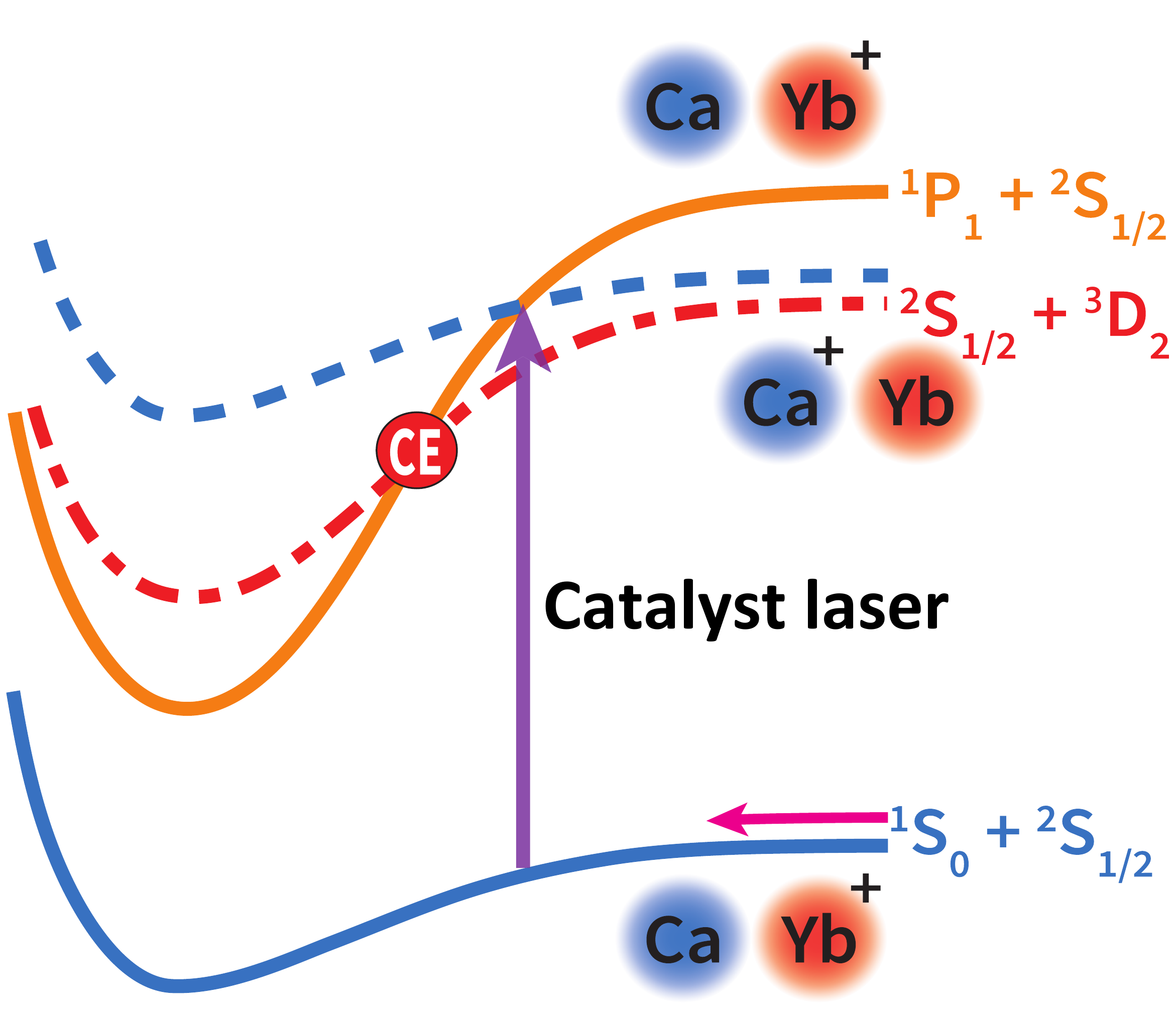}}

 \caption{\label{fig:molecularpots}\textbf{Long-range diabatic potential energy curves.} 
\textbf{(a)}  
Solid orange lines dissociate to the Ca(4s4p $^1$P$_1$) + Yb$^+$(6s $^2$S$_{1/2}$) limit. 
Three dash-dotted red lines dissociate to the Ca$^+$(4s $^2$S$_{1/2}$) + Yb(5d6s $^3$D$_2$) product limit. 
Finally, the dashed blue line corresponds to the dressed-state potential dissociating to Ca(4s$^2$ $^1$S$_0$) + Yb$^+$(6s $^2$S$_{1/2}$) plus one photon. 
Each curve corresponds to a total angular momentum and is labeled by one or more projection $|\Omega|$.
The two crossings between potentials relevant for the charge-exchange reactions are indicated with black circular
markers. 
The potential energy zero is located at the Ca(4s$^2$ $^1$S$_0$) + Yb$^+$(6s $^2$S$_{1/2}$) dissociation limit.
\textbf{(b)}
The first pathway corresponds to a collision between  an excited 4s4p~$^1$P$_1$ Ca atom with a ground-state 6s~$^2$S$_{1/2}$ Yb$^+$ ion.
The charge-exchange (CE) crossing is shown by a red circle.
The vertical wavy line represents spontaneous emission to the ground  Ca(4s$^2$ $^1$S$_0$) + Yb$^+$(6s $^2$S$_{1/2}$) channel, illustrating that spontaneous emission can occur before the atom-ion pair reaches short range.
\textbf{(c)}
The second pathway corresponds to a collision between  a ground-state 4s$^2$~$^1$S$_0$ Ca atom  with  a 6s~$^2$S$_{1/2}$ Yb$^+$ ion in the presence of a photon of the MOT laser.
The dashed blue curve corresponds to the dressed-state potential for this entrance channel. It has an avoided crossing with the excited Ca(4s4p~$^1$P$_1$) + Yb$^+$(6s~$^2$S$_{1/2}$) potential, and the coupling due to the MOT laser, shown as a purple arrow, can promote to the excited potential. 
Charge exchange then proceeds as in panel (b) if no spontaneous emission occurs.
\textbf{(d)}
In the presence of a catalyst laser, the incoming Ca(4s4p $^1$S$_0$) + Yb$^+$(6s $^2$S$_{1/2}$) potential is coupled to the reactive Ca(4s4p $^1$P$_1$) + Yb$^+$(6s $^2$S$_{1/2}$) potential at short range, where spontaneous emission is unlikely before reaching the charge-exchange crossings.
}

\end{figure*}

To understand this behavior, we first consider the typical case of charge exchange at low temperatures, where no measures are taken to overcome reaction blockading. Fig.~\ref{fig:molecularpots} shows the relevant CaYb$^+$ long-range diabatic potentials, labeled by the projection $\Omega$ of the total angular momentum onto the intermolecular axis, as a function of atom-ion separation $R$.
The entrance channel to the studied charge-exchange process, Ca(4s4p $^1$P$_1$) + Yb$^+$(6s $^2$S$_{1/2}$), has both a four-fold-degenerate ($|\Omega| = 1/2,3/2$) repulsive potential and two-fold-degenerate ($|\Omega| = 1/2$) attractive potential. 
Substantial non-radiative charge transfer only occurs to the Ca$^+$(4s $^2$S$_{1/2}$) + Yb(5d6s $^3$D$_{2}$) exit channel.
The Ca$^+$(4s $^2$S$_{1/2}$) + Yb(5d6s $^3$D$_{3}$) channel is energetically inaccessible to this entrance channel, and the Ca$^+$(4s $^2$S$_{1/2}$) + Yb(5d6s $^3$D$_{1}$) channel is only crossed at short range $R\approx 25$ $a_0$, where the the estimated couplings between these diabatic potentials, using the Heitler-London method~\cite{Tang98}, are too large to significantly contribute to the rate coefficient. 

Therefore, the non-radiative charge transfer is primarily driven by coupling of the $|\Omega| = 1/2$ entrance channel diabats at their crossings with the exit channels. 
This coupling arises from the molecular electrostatic interaction and therefore conserves $\Omega$, implying that only charge transfer to the $|\Omega| = 1/2$ exit channel diabats, at crossing points $R_c = 40.7$ $a_0$ and 42.3 $a_0$, is relevant. 
Since the electronic basis functions are very different for the two channels, the non-adiabatic coupling is localized and approximated by identical Lorentzians centered at each $R_c$.
The half width of this Lorentzian, $R_0$, is chosen to match the experimentally determined charge transfer rates.
In the absence of any additional means to overcome reaction blockading, the atom-ion pair can reach these crossing points and undergo a charge-exchange reaction via two pathways.
The first pathway is directly on the Ca(4s4p $^1$P$_1$) + Yb$^+$(6s $^2$S$_{1/2}$) entrance channel, where we determine the population of Ca atoms in the $^1$P$_1$ state by solving a rate equation derived from the optical Bloch equations, which includes the distance-dependent detuning of the MOT beams \cite{MMills17}. 
The second pathway describes a collision on the photon-dressed Ca(4s4p $^1$S$_0$) + Yb$^+$(6s $^2$S$_{1/2}$) potential, which is coupled to the Ca(4s4p $^1$P$_1$) + Yb$^+$(6s $^2$S$_{1/2}$)  state via the MOT laser. 
Because the MOT laser is tuned $2\pi\times34.6$~MHz below the asymptotic transition energy, as the atom-ion collide the laser is shifted into resonance.
At this point, there is a resonant amplification of the coupling to the Ca(4s4p $^1$P$_1$) + Yb$^+$(6s $^2$S$_{1/2}$) potential by the molecular interaction due to the large density of states near the threshold. 
This effect is calculated separately as this amplification is not included in the rate equation model, which only includes atomic states. 
%As such, this second pathway describes charge exchange with an entrance channel Ca(4s4p $^1$S$_0$) + Yb$^+$(6s $^2$S$_{1/2}$) coupled to the exit channel Ca$^+$(4s $^2$S$_{1/2}$) + Yb(5d6s $^3$D$_{2}$)  via the intermediate channel Ca(4s4p $^1$P$_1$) + Yb$^+$(6s $^2$S$_{1/2}$) .

Using the infinite-order sudden approximation (IOSA)~\cite{Pack74,Secrest75,Hunter75,Kouri79}, a coupled-channels calculation is performed on these potentials to determine the charge transfer cross-section, $\sigma(E,\ell)$.
The effect of spontaneous emission is included by classically computing the collision time on the entrance channel and determining the probability, $p(E,\ell)$, for a colliding pair to survive to $R_c$ without spontaneously emitting. 
The charge transfer rate constant is then determined as
\begin{align}
k = \sum_{\ell = 0}^{\infty}(2\ell+1)p(E,\ell)\sigma(E,\ell)
\end{align}
where $\ell$ is the average orbital angular momentum quantum number used in the IOSA and $E$ is the collision energy. The resulting rate constant is displayed alongside the data in Fig.~\ref{fig:trap}(d) for $R_0 = 0.39$ $a_0$, with and without the inclusion of $p(E,\ell)$. A detailed description of the excited-state potentials and charge transfer can be found in \cite{Ming2018}.

\begin{figure}[t]
 \centering
\includegraphics[width=0.9\textwidth]{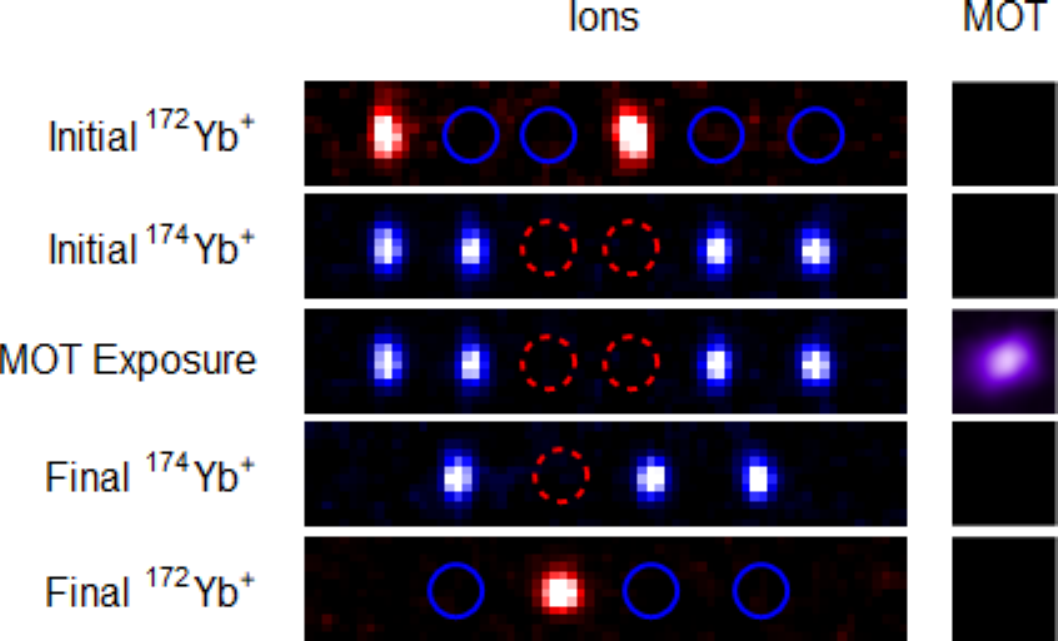}
 \caption{\label{fig:cexexp} \textbf{Dual-isotope technique.} False-color fluorescence images of the Yb$^+$ ions and the Ca MOT (not to scale) illustrating the dual-isotope method used to measure the low decay rate of $^{172}$Yb$^+$($^2$S$_{1/2}$). 
In a typical experimental sequence, we first trap $^{172}$Yb$^+$ and $^{174}$Yb$^+$, while laser-cooling only $^{172}$Yb$^+$ ions (shown in red), while $^{174}$Yb$^+$ ions (shown as blue circles) remain dark.
We then switch the 369 nm cooling laser frequency to cool $^{174}$Yb$^+$ ions (shown in blue), while the $^{172}$Yb$^+$ ions (shown as red dashed circles) remain dark.
To prevent build-up of $^{172}$Yb$^+$ ions in the $^2$D$_{3/2}$ state, both isotopes have a 935 nm repump laser present at all times.
We then overlap the MOT with the laser-cooled $^{174}$Yb$^+$ ions as well as the ground-state $^{172}$Yb$^+$($^2$S$_{1/2}$) ions for a variable amount of time.
Finally, we switch the cooling laser frequency once again to cool $^{172}$Yb$^+$ ions and use fluorescence to measure the final number of $^{172}$Yb$^+$ ions.
}

\end{figure}

\begin{figure}[t!]
 \centering
 \hfill
 \sidesubfloat[]{\includegraphics[width=.95\textwidth]{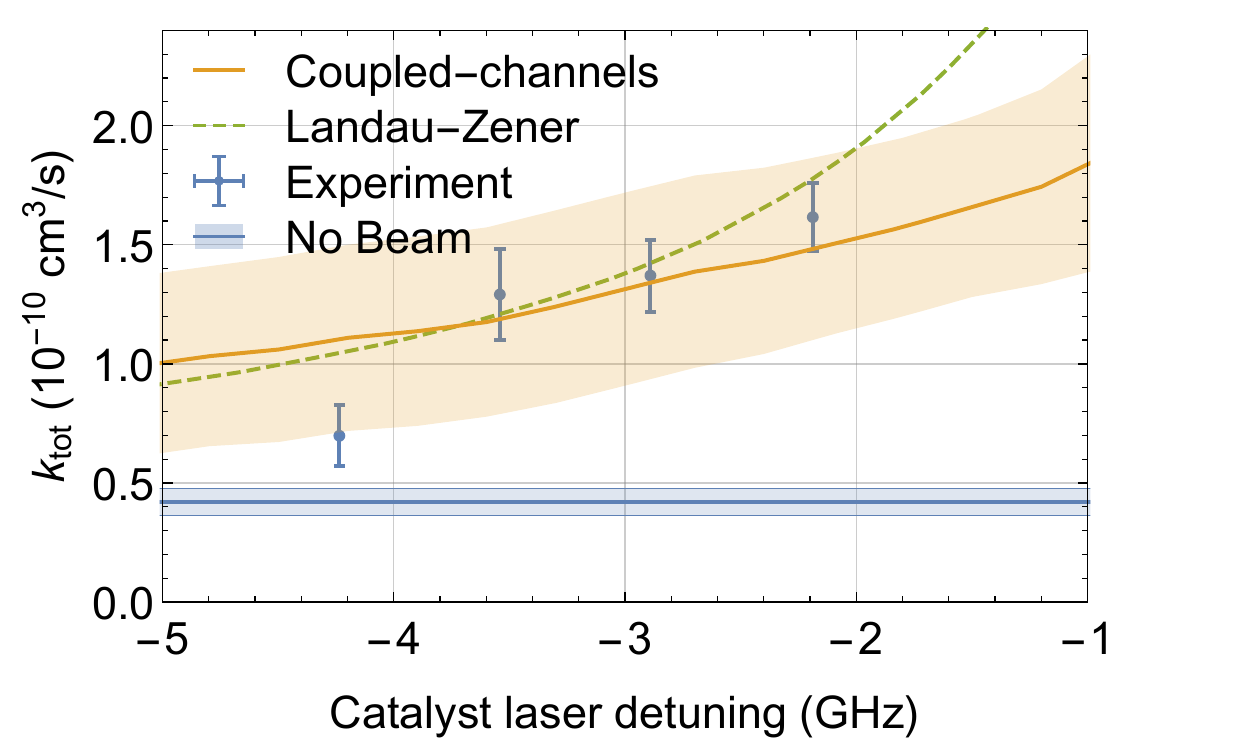}\label{antishielding:sub1}}
 \hfill
 \sidesubfloat[]{\includegraphics[width=.95\textwidth]{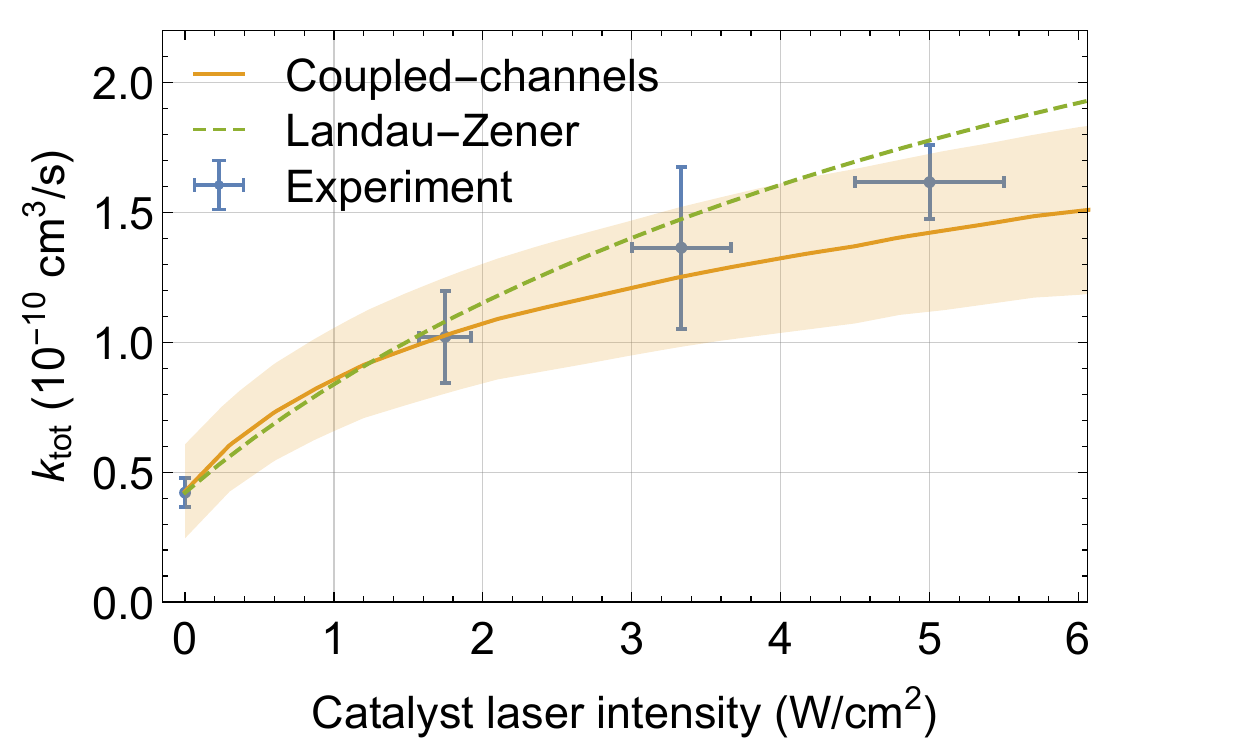}\label{antishielding:sub2}}
 
 \caption{\textbf{Removing suppression with addition of a catalyst laser.} Total charge-exchange rate coefficient as a function of catalyst laser \textbf{(a)} frequency and \textbf{(b)} intensity. Plotted alongside experimental data are the results of a coupled-channels calculation and an estimate using the Landau-Zener approximation. For reference, the experimental rate with no catalyst beam is shown. Error bars correspond to the standard error in experimental measurements and error bands include uncertainties from the theoretical simulations and experimental parameters. Horizontal error bars in (a) are smaller than the plot marker. Details are found in the Supplementary Material.}\label{fig:antishielding}
\end{figure}

Given that this reaction blockading is expected to occur in all low-temperature excited-state collisions, it is desirable to develop a method to remove it.
Here, we demonstrate one such means for removing the reaction blockading effect.
Building on ideas developed for control of hyperfine-changing collisions~\cite{Gould95a,Gould95,PA}, we apply a strong laser, dubbed the catalyst laser, that couples the ground-state potential with an excited state at short range.
This allows selection of the excited-state reaction channel and may, in principle, be used to select a desired reaction product in polyatomic systems.

The operation of the technique is sketched in Fig.~\ref{fig:molecularpots}(d), where the CaYb$^+$ molecular potentials dressed by the photon energy of the applied laser are shown.
Near the catalyst laser avoided crossing distance $R_{CL}$, the catalyst beam couples the upper and lower potentials, promoting the complex to the Ca($^1$P$_1$) + Yb$^+$($^2$S$_{1/2}$) potential at short range.
The probability of promotion can be estimated from Landau-Zenter transition theory as $P(\Omega_R) = 1 - \textrm{Exp}\left[-\pi (\hbar \Omega_R)^2 / \left(2 \hbar v \frac{\partial{}}{\partial{R}}\Delta E\right)\right]$, where $\Omega_R$ is the Rabi frequency of the catalyst beam, $v$ is the radial velocity, and $\Delta E$ is the energy difference between the diabatic potentials \cite{Wittig2005}.
Thus, for a scattering event with rate constant, $k$, this technique yields an experimentally observable rate $k_o = k P(\Omega_R)e^{-\Delta t(R_{CL})/\tau_{\textrm{P}}}$, where $\Delta t(R_{CL})$ is the time required for the atom-ion pair to propagate from $R_{CL}$ to short range and $\tau_{\textrm{P}}$ is the lifetime of the excited potential.

In order to test the catalyst laser technique at the lowest possible collision energy, where the suppression is strongest, the ions cannot be shuttled but must be arranged in a stationary linear ion chain overlapped with the MOT.
Due to collisional heating effects~\cite{Chen2013,Chen2014,Schowalter2016}, the linear ion chain cannot be maintained during MOT exposure without active laser cooling.
However, if the ions are laser cooled, the charge-exchange rates include collisions originating in the Yb$^+$($^2$P$_{1/2}$) and Yb$^+$($^2$D$_{3/2}$) states~\cite{Rellergert2011}.
Therefore, in order to isolate the Ca($^1$P$_1$) + Yb$^+$($^2$S$_{1/2}$) charge-exchange rate without the shuttling method, we develop and implement a dual-isotope technique (see Fig.~\ref{fig:cexexp}) for collision rate measurement.
Specifically, we simultaneously trap both $^{172}$Yb$^+$ and $^{174}$Yb$^+$ ions, while laser-cooling only the $^{174}$Yb$^+$ ions, which, in turn, sympathetically cool the $^{172}$Yb$^+$ ions.
As the $^{172}$Yb$^+$ ions are only sympathetically cooled, they remain in the 6s $^2$S$_{1/2}$ state and do not experience the large charge-exchange rate of Yb$^+$ ions in the 6p $^2$P$_{1/2}$ state.
Due to off-resonant scattering of the cooling laser for the $^{174}$Yb$^+$ ions, it is necessary to apply a repumping laser for the $^{172}$Yb$^+$ ions to prevent population from accumulating in the 5d $^2$D$_{3/2}$ state.
Therefore, by monitoring the number of $^{172}$Yb$^+$ ions with time, we isolate and measure the charge exchange of Ca($^1$P$_1$) + Yb$^+$($^2$S$_{1/2}$).

Figs.~\ref{fig:antishielding}(a) and (b) show the results of using this dual-isotope technique to monitor Ca($^1$P$_1$) + Yb$^+$($^2$S$_{1/2}$) charge-exchange reactions as a function of the detuning and intensity of the catalyst laser, respectively, at a collision temperature of $\sim$50 mK.
For large detunings, although the atom-ion pair is promoted at a small value of $R_{CL}$, increasing the likelihood of reaching short range before spontaneous emission, the large value of $\frac{\partial}{\partial R} \Delta E$ and the high velocity of the reactants leads to a lower probability of promotion to the reactive potential from Landau-Zener transition theory.
For the given experimental intensity of 5~W/cm\textsuperscript{2}, the catalyst beam cannot be closer to resonance than $\sim -2$ GHz due to adverse effects on the MOT.

The dependence of the measured rate on the catalyst laser intensity can be understood by the increased probability of promotion to the reactive potential given by Landau-Zener transition theory for increasing Rabi frequencies $\Omega_R$.
Also shown are the results of a coupled-channels calculation.
Here, the rates are calculated by allowing for, in addition to the two previously discussed pathways from the MOT laser, a catalyst-laser-enhanced charge-exchange pathway, coupling the Ca($^1$S$_0$) + Yb$^+$($^2$S$_{1/2}$) entrance channel to the Ca$^+$($^2$S$_{1/2}$) + Yb($^3$D$_2$) exit channel via the intermediate Ca($^1$P$_1$) + Yb$^+$($^2$S$_{1/2}$) channel.
The experimental data shows good agreement with both the coupled-channels calculations and the simple Landau-Zener model, supporting this interpretation of the results.

In summary, we have investigated and engineered electronically excited-state collisions of Ca with Yb$^+$ at low collision energy.
Using a method for precise control of collision energy, we find that the interaction of the atom with the ion leads to a strong shift between the ground and excited atomic states, causing any laser addressing the bare atomic transition frequency to be shifted from resonance, even at long range.
Thus, at low collision energy, an atomic excited state is likely to undergo spontaneous emission before reaching short range. This leads to a strong suppression of scattering events that occur on any potential corresponding to an atomic excited state.
These features are expected to be universal at low temperature for systems with short-lived electronic excitations and long-ranged interactions.
To overcome this suppression, we demonstrate a technique using a catalyst laser, which selectively excites colliding molecular complexes at short range.
This technique removes the observed reaction blockading, allowing excited-state collisions to be studied even at ultracold temperatures.
As low-temperature techniques provide precise information about the underlying dynamics, this technique should find use as a general tool for studying excited-state collisions.
Further, because the technique selectively excites the colliding pair to a chosen potential it may be used as a means to select a desired product outcome in polyatomic chemical reactions.

Finally, the reaction blockading effect observed and controlled here is extremely important for the growing field of hybrid atom-ion trapping, where sympathetic cooling of ions with laser-cooled atoms is being pursued~\cite{Rellergert2013,Hudson2016}.
The existence of this reaction blockading effect means that detrimental chemical reactions from excited atomic states, which are energetically unavoidable, will not occur during the sympathetic cooling process.
Thus, a large variety of molecular ions can be cooled by laser-cooled atoms without loss to unwanted chemical reactions.

We thank Wesley Campbell for helpful discussions.
This work was supported in part by National Science Foundation (PHY-1255526, PHY-1415560, and DGE-1650604) and Army Research Office (W911NF-15-1-0121, W911NF-14-1-0378, and W911NF-13-1-0213) grants.
Work at the Temple University is supported by the MURI Army Research Office Grant No. W911NF-14-1-0378 and Grant No. W911NF-17-1-0563, the U.S. Air Force Office of Scientific Research Grant No. FA9550-14-1-0321 and the NSF Grant No. PHY-1619788.

\bibliography{mybib}

\end{document}